\input{aipcheck}

\documentclass[
    ,final            
  ]
  {aipproc}

\layoutstyle{6x9}

\usepackage{epsfig}

\begin{document}

\title{Hard core attraction in hadron scattering and 
the family of the Ds meson molecule}
\author{Pedro Bicudo}{address={
Dep. F\'{\i}sica and CFIF, Instituto Superior T\'ecnico, Av.
Rovisco Pais
1049-001 Lisboa, Portugal}}
\author{Gon\c{c}alo M. Marques}{address={
Dep. F\'{\i}sica and CFIF, Instituto Superior T\'ecnico, Av.
Rovisco Pais
1049-001 Lisboa, Portugal}}
\begin{abstract}
We study the discovered Ds(2317) at BABAR, CLEO and BELLE, and find that it 
belongs to a class of strange multiquarks, which is equivalent to the class of 
kaonic molecules bound by hard core attraction. In this class of hadrons a kaon 
is trapped by a s-wave meson or baryon.
To describe this class of multiquarks we apply the Resonating Group Method, and 
extract the hard core kaon-meson(baryon)interactions. We derive a criterion to classify 
the attractive channels. We find that the mesons f0(980), Ds(2457), Bs scalar and axial, and 
also the  baryons with the quantum numbers of $\Lambda, \Xi_c, \Xi_b$ and also 
$\Omega_{cc}, \Omega_{cb}$ and $\Omega_{bb}$ belong to the new hadronic class of the Ds(2317).
\end{abstract}
\maketitle


\par
Recently new narrow scalar resonances $D_s(2317)$ and $D_s(2457)$ were discovered 
at BABAR  
\cite{Babar}, 
CLEO
\cite{Cleo}
and BELLE
\cite{Belle}.
These positive parity resonances 
\cite{Rupp2}
were also predicted by Nowak, Ro and Zahed 
\cite{Nowak,Bardeen}
as chiral partners of the well known negative parity $D_s(1968)$ and $D_s(2112)$.
Here we find that the new $D_s$ resonances can be undesrtood as tetraquarks, 
or equivalently as $D$-$K$ molecules 
\cite{Tornqvist}
bound by the hardcore attraction.
These are not standard hadrons because they are neither quark-antiquark mesons nor three 
quark baryons. 
The experimental discovery of these hadrons also prompts us to study the 
new hadronic class which includes the $D_s(2317)$. 
In our framework the masses and couplings of hadrons are microscopically computed
at the quark level and in a chiral invariant framework,
\cite{Bicudo3,Bicudo1,Bicudo2}.

\par
In this talk we study the family of all possible narrow tetraquark 
and pentaquark resonances 
\cite{Jaffe}
where the quark $s$, or the Kaon play a crucial role. 
We start by reviewing the Resonating Group Method (RGM) 
\cite{Wheeler}. 
The RGM, together with chiral symmetry, produces hard core 
hadron-hadron potentials, which can be either repulsive or attractive. 
We derive a criterion to discriminate which systems
bind and which are unbound. We apply this criterion to find, among the s-wave 
hadrons, the candidates to trap a kaon. Finally we compute the binding energy
of the selected hadrons, the positive parity mesons $f_0(980),  
D_s^{(0+)}, \, D_s^{(1+)}, \, B_s^{(0+)}, \, B_s^{(1+)}$ and the
negative parity baryons $\Lambda(1405), \Xi_c^{(-)}, \Xi_b^{(-)}$ and also $\Omega_{cc}^{(-)}, \Omega_{cb}^{(-)}$
and $\Omega_{bb}^{(-)}$.

\par
The RGM computes the effective multiquark energy 
using the matrix elements of the microscopic quark-quark interactions. 
The multiquark state is decomposed in anti-symmetrized combinations 
of simpler colour singlets, the baryons and mesons. 
The RGM was first used in hadronic physics by Ribeiro 
\cite{Ribeiro} 
to show that in exotic
hadron-hadron scattering, the quark-quark potential together with
the Pauli repulsion of quarks produces a repulsive short range
interaction. For instance this explains the $N - N$ hard core
repulsion, preventing nuclear collapse. Deus and Ribeiro
\cite{Deus} 
used the same RGM to show that, in non-exotic
channels, the quark-antiquark annihilation could produce a short
core attraction. This is confirmed in several studies
\cite{Bicudo3,Bicudo1,Bicudo2,Ribeiro,Deus,Bicudo}.

\par
\begin{figure}[t]
\epsfig{file=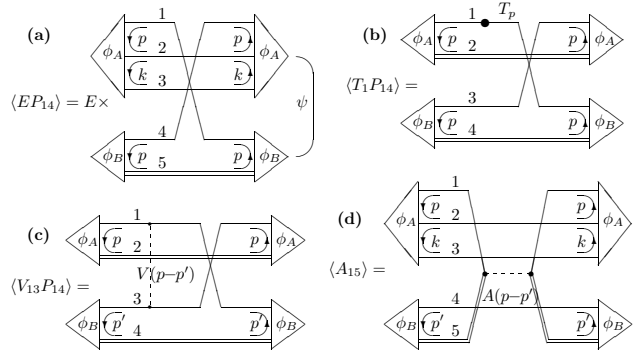,width=8.5cm} \caption{We show examples of
RGM overlaps, in (a) the norm overlap for the meson-baryon
interaction, in (b) a kinetic overlap the meson-meson interaction,
in (c) an interaction overlap the meson-meson interaction, in (d)
the annihilation overlap for the meson-baryon interaction.}
\label{RGM overlaps}
\end{figure}

The energy of the multiquark is computed with the matrix elements
of the hamiltonian,
\begin{equation}
H=\sum_i T_i +\sum_{i<j} V_{ij} +\sum_{i \bar j} A_{i \bar j} \ ,
\end{equation}
which includes the quark(antiquark) kinetic energies, the 
quark(antiquark)-quark(antiquark) interaction proportional
to the colour dependent 
${ \vec \lambda_i \over 2} \cdot { \vec \lambda_j \over 2}$
, and the 
quark-antiquark annihilation. 
Once the internal energy of each cluster is accounted, 
and we also use the antisymmetry of the baryon wave-function,
the remaining energy of the meson-meson or baryon-meson system is 
computed with the overlap,
\begin{eqnarray}  
{\cal O}_{bar \, A \atop mes \, B}&=&
3 \langle \phi_B \, \phi_A | -( V_{14}+V_{15}+2V_{24}+2V_{25} )P_{14} 
\nonumber \\
&& +A_{15} | \phi_A \phi_B \rangle
\nonumber \\
{\cal O}_{mes \, A \atop mes \, B}&=&
\langle \phi_B \, \phi_A | (1+P_{AB})[ -( V_{13}+V_{23}+V_{14}+V_{24})
\nonumber \\
&& \times P_{13} + A_{23} + A_{14} ] | \phi_A \phi_B \rangle
\ ,
\label{overlap kernel} 
\end{eqnarray}
where we use an obvious notation (see Fig. \ref{RGM overlaps}).
The colour, spin and flavour contributions to the overlaps simply provide
an algebraic factor. The momentum (or space)
integrations, is estimate them with the
variational method to produce a geometrical overlap,
\begin{equation}
\sum_{sfc} \int_p \phi_A^\dagger \phi_B^\dagger H \phi_B \phi_A
\simeq M \sum_{sfc} \int_p \phi_A^\dagger \phi_B^\dagger \phi_B \phi_A \ ,
\label{variational}
\end{equation}
where $M$ is the sum of hadronic masses corresponding to the
Hamiltonian $H$. 
%
%

\par
When spontaneous chiral symmetry breaking is
included in the quark model 
\cite{Bicudo3,Bicudo4} 
the annihilation potential turns ou to be crucial 
\cite{Bicudo1,Bicudo2} 
to understand the low mass of the $\pi$. The annihilation potential
$A$ is also present in the $\pi$ Salpeter equation where it
cancels most of the kinetic energy and confining potential $2T+V$.
From the quark model with chiral symmetry breaking we get the sum
rules \cite{Bicudo3},
\begin{eqnarray}
\langle 2 T+V +A \rangle_{S=0} &=& M_\pi \simeq 0
\nonumber \\
\langle 2T+V \rangle_{S=0} &=& {2 \over 3} (2M_N-M_\Delta)
\nonumber \\
\Rightarrow \langle A \rangle_{S=0} &\simeq& - {2 \over 3} (2M_N-M_\Delta) 
\ .
\label{sum rules}
\end{eqnarray}
In eq. (\ref{sum rules}) 
the matrix elements are evaluated for s-wave and
spin 0 wave-functions of light quarks. Similar sum rules also
exist for spin 1 wave-functions, involving only the mass of the
$\Delta$ baryon. The annihilation potential
for the strange quark mass is smaller by a factor of $\sigma$ which is a 
power of the constituent quark mass ratio $M_{u,d} /M_s$. 
From eqs. (\ref{sum rules}) it is clear that the annihilation
potential provides an attractive (negative) overlap. 

\par
In what concerns the quark-quark(antiquark) potential,
in the present case of s-wave states, with a $S=0$ kaon, the 
only potential which may contribute is the hyperfine potential, 
proportional to 
${ \vec \lambda_i \over 2} \cdot { \vec \lambda_j \over 2}
\ \vec S_i  \cdot  \vec S_j $. 
We find that the total matrix element is a hyperfine splitting,
\begin{equation}
\langle P_{13}(V_{13}+V_{23}+V_{14}+V_{24})\rangle =4 {2 \over 3}(M_\Delta-M_N) 
\end{equation}
and it is repulsive (positive).

\par
These results are
independent of the particular quark model that we choose to
consider, providing it is chiral invariant.
We therefore arrive at the criterion 
\\
- {\em whenever the two interacting hadrons have
a common flavour, the repulsion is increased,
\\
- when the two interacting hadrons have a matching quark and antiquark the attraction 
is enhanced }.

\par
In this paper we are interested in the class of resonances which can be 
understood as a S=-1 kaon $s \bar u$ or $s \bar d$ trapped by a s-wave hadron. 
With our criterion we can exclude all hadrons with an antiquark $\bar u $ or $\bar d$ 
or with a quark $s$ 
because the exchange overlap $\langle P_{13}\rangle$ would be allowed, and this
certainly contributes to repulsion. We assume that the attraction is
not sufficient to overcome this repulsion. In what concerns attraction
we need a quark $u$ or $d$ in the s-wave partner of the kaon, in order
to produce annihilation. 
This excludes the mesons $\eta, \, \eta' \, \omega, \, \phi$. 
Moreover we specialize in systems which are possible to study experimentally,
where the kaon partner is a hadronic resonance with a very narrow width.
This restricts the kaon partner to s-wave mesons and baryons,
and excludes wider resonances like the meson $\rho$ and the baryon $\Delta$.
The pion is also excluded because it is too light to bind to the kaon , all 
that we can get is a very broad resonance, the kappa resonance \cite{Rupp1}. 
Therefore the hadrons which are best candidates to strongly bind the Kaons
$s \bar l$ are the s-wave hadrons with flavour 
$l\bar s , \, l \bar c, \, l \bar b, \, lll, \, llc, \, llb , \, lcc, \, lcb, \, lbb$. 
This is expected to result in the $f_0(980)$ 
\cite{Rupp1}, 
the $D_s(2320$ 
\cite{Rupp2} 
the $D_s(2463)$
\cite{Babar}, 
the $\Lambda(1405)$ and several other predicted resonances.

\par
A convenient basis for the meson and baryon wave-functions is the 
harmonic oscillator basis, parametrized by the inverse radius $\alpha$. We summarize
\cite{Bicudo1,Bicudo2,Bicudo} the effective potentials computed
for the different channels,
\begin{eqnarray}
V_{K-K}&=& 2 (1+\sigma)^2{1\over 6} \langle A \rangle \,  |\phi_{000}^\alpha \rangle \langle \phi_{000}^\alpha| \ ,
\nonumber \\
V_{K-D,D^*,B,B^*}&=& 2 {1 \over 6} \langle A \rangle \, |\phi_{000}^\alpha \rangle \langle \phi_{000}^\alpha| \ , 
\nonumber \\
V_{K-N}&=& 4 {1 \over 6} \langle A \rangle \, |\phi_{000}^\alpha \rangle \langle \phi_{000}^\alpha| \ ,
\nonumber \\
V_{K-\Sigma_c,\Sigma_b}&=& 
{7 \over 3} {1 \over 6} \langle A \rangle \, |\phi_{000}^\alpha \rangle \langle \phi_{000}^\alpha| \ ,
\nonumber \\
V_{K-\Xi_{cc},\Xi_{cb},\Xi_{bb}}&=& 2 {1 \over 6} \langle A \rangle \, 
|\phi_{000}^\alpha \rangle \langle \phi_{000}^\alpha| \ , 
\label{zero p}
\end{eqnarray}
where the colour and spin factors contribute respectively with $1/3$ and $1/2$, 
$\langle A \rangle$ is of the order of 430 MeV and the geometrical factor provides
the separable potential $|\phi_{000}^\alpha \rangle \langle \phi_{000}^\alpha|$ .
The remaining factor is the flavour factor, the only one that turns out to 
differ in the s-wave kaon-hadron annihilation. The parameter $\alpha$ is
are the only model one, and from the $D_s(2320)$ channel we $\alpha$=285 MeV. 

\par
This parametrization hard core interaction in a separable potential,
enables us to use standard techniques
\cite{Bicudo} 
to exactly compute the scattering $T$ matrix.
The binding occurs when the $T$ matrix has a pole for
a negative relative energy, and this happens when,
\begin{equation}
- 4 \mu v \ge { \alpha^3 \over \beta } \ .
\end{equation}

\par
The results are displayed in Tables \ref{binding energies} and \ref{excessive binding}. 
We conclude that the
$f_0(980$ and the $D_s^{(0+)}, \, D_s^{(1+)}, \, B_s^{(0+)}, \, B_s^{(1+)}$
belong to the same class of tetraquark hadronic resonances. This
class is consistent with the picture of a kaon trapped by a
s-wave meson. In what concerns pentaquarks, where the kaon is
trapped for instance by a nucleon to produce a $\Lambda$, or by
other hadrons with u or d light quarks, our 
results predict that we have also have binding with the quantum
numbers of the $\Lambda$ and $\Xi_c^+$, $\Xi_c^0$, $\Xi_b^0$, $\Xi_b^-$
and also of the $\Omega_{cc}$, $\Omega_{cb}$, and $\Omega_{bb}$. 

\par
Here we neglected the coupling to the pion-hadron channels. This is correct
for the multiquarks of Table \ref{binding energies} where this coupling is
isospin violating. However in the systems of Table \ref{excessive binding}, 
where we found a deep binding, we plan to include the coupling to the 
pion-hadron channels. This coupling is expected to decrease substantially the
binding energy. We also expect that our method addresses the 
protonomium recently discovered at BES
\cite{BES} 
and the deuson X(3872) recently discovered at BELLE
\cite{BELLE2}.

\begin{theacknowledgments}
We are very grateful to Emilio Ribeiro for discussions on the
RGM and to George Rupp for discussions on
hadronic resonances. 
The work of G. Marques is supported by Funda\c c\~ao para a
Ci\^encia e a Tecnologia under the grant SFRH/BD/984/2000.
%
%
\begin{table}[t]
\begin{tabular}{c|cccccc}
channel                   
& $\mu_{exp}$  & $v_{th}$&$\alpha=\beta$& $B_{th}$& $B_{exp}$ \\
\hline 
$ D_s(2317)= {K^- {\bar D}^0+{\bar K}^0 D^- \over \sqrt{2}  }   $ 
& 392 & -143 &  {\em 285 } &  {\em 46}& 46
 \\
$  D_s(2457)= {K^- {\bar D}^{*0}+{\bar K}^0 D^{*-} \over \sqrt{2} }    $ 
& 398 & -143 &  285 & 47 & 46
 \\
$  K^- {\bar B}^0+{\bar K}^0 B^- \over \sqrt{2}     $ 
& 453 & -143 &  285 &  55 & -
 \\
$ K^- {\bar B}^{*0}+{\bar K}^0 B^{*-} \over \sqrt{2}     $ 
& 454 & -143 &  285 & 55 & -
 \\
$  { K^- \Xi^{++}_{cc} + {\bar K}^0 \Xi^+_{cc} \over \sqrt{2}  }    $ 
& 442 & -143 &  285 & 53 & -
 \\
$  { K^- \Xi^{+}_{cb}  + {\bar K}^0 \Xi^{0}_{cb} \over \sqrt{2}  }    $ 
& 466 & -143 &  285 & 56 & -
 \\
$  { K^- \Xi^{0}_{bb}  + {\bar K}^0 \Xi^{-}_{bb} \over \sqrt{2}  }    $ 
& 475 & -143 &  285 & 58 & -
 \\
\hline
\end{tabular}
\caption{ This table summarizes the parameters $\mu , \, v \,
,\alpha \, , \beta$ 
 and binding energies $B$  (in MeV).
The italic binding energy $B_{th}$  of the $D_s(1327)$ is fitted from experiment}
\label{binding energies}
\end{table}
%
%
%
%
%
\begin{table}[t]
\begin{tabular}{c|cccccc}
channel                   
& $\mu_{exp}$  & $v_{th}$&$\alpha=\beta$& $B_{th}$& $B_{exp}$ \\
\hline 
$ f_0(980)= {K^- K^+ + {\bar K}^0 K^0 \over \sqrt{2} } $ 
& 248 & -207&  285 & 59 & 12 $\pm$ 10
\\
$ \Lambda(1405)= { K^- p + {\bar K}^0 n\over \sqrt{2}  }    $ 
& 325 & -286 &  285 & 149 & 30 $\pm$ 4 
 \\
$  { K^- \Sigma^{++}_c + {\bar K}^0 \Sigma^{+}_c \over \sqrt{2}  }    $ 
& 412 & -167 &  285 & 68 & -
 \\
$  { K^- \Sigma^{+}_b + {\bar K}^0 \Sigma^{0}_b \over \sqrt{2}  }    $ 
& 456 & -167 &  285 & 75 & -
 \\
\hline
\end{tabular}
\caption{ 
Results for channels open to pion decay.
}
\label{excessive binding}
\end{table}

\end{theacknowledgments}


\end{document}